\documentclass[RNAAS]{aastex63}

\newcommand{\niceurl}[1]{\href{#1}{#1}}

\newcommand\gaia{{\it Gaia}}

\begin{document}

\title{Preliminary Target Selection for the DESI Milky Way Survey (MWS)}


\author[0000-0002-0084-572X]{Carlos Allende Prieto}
\affiliation{Instituto de Astrof\'{i}sica de Canarias, V\'{i}a L\'actea S/N, 38205 La Laguna, Tenerife, Spain}
\affiliation{Universidad de La Laguna, Departamento de Astrof\'{\i}sica, 38206 La Laguna, Tenerife, Spain}
\author[0000-0001-8274-158X]{Andrew P. Cooper}
\affiliation{Institute of Astronomy and Department of Physics, National Tsing Hua University, 101 Kuang-Fu Rd.\ Sec.\ 2, Hsinchu 30013, Taiwan}
\author{Arjun Dey} 
\affiliation{NSF's NOIRLab, 950 N. Cherry Ave., Tucson, AZ 85719, USA}
\author[0000-0002-2761-3005]{Boris T. G\"ansicke} 
\affiliation{Department of Physics, University of Warwick, Coventry, CV4 7AL, UK}
\author[0000-0003-2644-135X]{Sergey E. Koposov}
\affiliation{Institute for Astronomy, University of Edinburgh, Royal Observatory, Blackford Hill, Edinburgh EH9 3HJ, UK}
\affiliation{McWilliams Center for Cosmology, Carnegie Mellon University, 5000 Forbes Ave, PA, 15213, USA}
\affiliation{Institute of Astronomy, University of Cambridge, Madingley Road, Cambridge CB3 0HA, UK}
\author[0000-0002-9110-6163]{Ting Li} 
\altaffiliation{NHFP Einstein Fellow}
\affiliation{Observatories of the Carnegie Institution for Science, 813 Santa Barbara St., Pasadena, CA 91101, USA}
\affiliation{Department of Astrophysical Sciences, Princeton University, Princeton, NJ 08544, USA}
\author{Christopher Manser}
\affiliation{Department of Physics, University of Warwick, Coventry, CV4 7AL, UK}
\author{David L. Nidever}
\affiliation{NSF's NOIRLab, 950 N. Cherry Ave., Tucson, AZ 85719, USA}
\affiliation{Department of Physics, Montana State University, P.O. Box 173840, Bozeman, MT 59717-3840}
\author{Constance Rockosi}
\affiliation{Department of Astronomy and Astrophysics, University of California Santa Cruz, 1156 High Street, Santa Cruz, CA, 95064, USA}
\affiliation{University of California Observatories}
\author{Mei-Yu Wang}
\affiliation{McWilliams Center for Cosmology, Carnegie Mellon University, 5000 Forbes Ave, PA, 15213, USA}
\affiliation{Department of Physics, Carnegie Mellon University, Pittsburgh, Pennsylvania 15312, USA}
\author[0000-0001-5200-3973]{David S. Aguado}
\affiliation{Institute of Astronomy, University of Cambridge, Madingley Road, Cambridge CB3 0HA, UK}
\author[0000-0002-8622-4237]{Robert Blum}
\affiliation{Vera C. Rubin Observatory/NSF's NOIRLab, 950 N. Cherry Ave., Tucson, AZ, 85719, USA}
\author{David Brooks}
\affiliation{Department of Physics \& Astronomy, University College London, Gower Street, London, WC1E 6BT, UK}
\author{Daniel J. Eisenstein}
\affiliation{Harvard-Smithsonian Center for Astrophysics, 60 Garden St., Cambridge, MA 02138, USA}
\author[0000-0002-2611-0895]{Yutong Duan}
\affiliation{Physics Department, Boston University, 590 Commonwealth Avenue, Boston, MA 02215, USA}
\author{Sarah Eftekharzadeh}
\affiliation{Department of Physics and Astronomy, The University of Utah, 115 South 1400 East, Salt Lake City, UT 84112, USA}
\author[0000-0001-9632-0815]{Enrique Gazta\~naga}
\affiliation{Institute of Space Sciences (ICE, CSIC), 08193 Barcelona, Spain}
\affiliation{Institut d\' ~Estudis Espacials de Catalunya (IEEC), 08034 Barcelona, Spain}
\author{Robert Kehoe}
\affiliation{Department of Physics, Southern Methodist University, 3215 Daniel Avenue, Dallas, TX 75275, USA}
\author{Martin Landriau}
\affiliation{Lawrence Berkeley National Laboratory, 1 Cyclotron Road, Berkeley, CA 94720, USA}
\author[0000-0003-1700-5740]{Chien-Hsiu Lee}
\affiliation{NSF's NOIRLab, 950 N. Cherry Ave., Tucson, AZ 85719, USA}
\author[0000-0003-1887-1018]{Michael E. Levi}
\affiliation{Lawrence Berkeley National Laboratory, 1 Cyclotron Road, Berkeley, CA 94720, USA}
\author[0000-0002-1125-7384]{Aaron M. Meisner}
\affiliation{NSF's NOIRLab, 950 N. Cherry Ave., Tucson, AZ 85719, USA}
\author{Adam D. Myers}
\affiliation{University of Wyoming, 1000 E. University Ave., Laramie, WY 82071, USA}
\author{Joan Najita}
\affiliation{NSF's NOIRLab, 950 N. Cherry Ave., Tucson, AZ 85719, USA}
\author{Knut Olsen}
\affiliation{Community Science and Data Center/NSF's NOIRLab, 950 N. Cherry Ave., Tucson, AZ, 85719, USA}
\author{Nathalie Palanque-Delabrouille}
\affiliation{IRFU, CEA, Universit\'e Paris-Saclay, F-91191 Gif-sur-Yvette, France}
\author{Claire Poppett}
\affiliation{Lawrence Berkeley National Laboratory, 1 Cyclotron Road, Berkeley, CA 94720, USA}
\affiliation{Space Sciences Laboratory, University of California, Berkeley, 7 Gauss Way, Berkeley, CA  94720, USA}
\author{Francisco Prada}
\affiliation{Instituto de Astrofisica de Andaluc\'{i}a, Glorieta de la Astronom\'{i}a, s/n, E-18008 Granada, Spain}
\author[0000-0002-5042-5088]{David J. Schlegel}
\affiliation{Lawrence Berkeley National Laboratory, 1 Cyclotron Road, Berkeley, CA 94720, USA}
\author{Michael Schubnell}
\affiliation{Department of Physics, University of Michigan, Ann Arbor, MI 48109, USA}
\author{Gregory Tarl\'e}
\affiliation{Department of Physics, University of Michigan, Ann Arbor, MI 48109, USA}
\author[0000-0002-6257-2341]{Monica Valluri}
\affiliation{Department of Astronomy, University of Michigan, Ann Arbor, MI 48109, USA}
\author[0000-0003-2229-011X]{Risa~H.~Wechsler}
\affiliation{Kavli Institute for Particle Astrophysics and Cosmology and Department of Physics, Stanford University, Stanford, CA 94305, USA}
\affiliation{SLAC National Accelerator Laboratory, Menlo Park, CA 94025, USA}
\author{Christophe Y\`eche}
\affiliation{IRFU, CEA, Universit\'e Paris-Saclay, F-91191 Gif-sur-Yvette, France}


\begin{abstract}

The DESI Milky Way Survey (MWS) will observe $\ge$8 million stars between $16<r<19$~mag, supplemented by observations of brighter targets under poor observing conditions. 
The 
survey will permit an accurate determination of stellar kinematics and population gradients; characterize diffuse substructure in the thick disk and stellar halo; enable the discovery of extremely metal-poor stars and other rare stellar types; and improve constraints on the Galaxy's 3D dark matter 
distribution 
from halo star kinematics. 
MWS will also enable a detailed characterization of the stellar populations within 100~pc of the Sun, including a complete census of white dwarfs. The target catalog from the preliminary selection described 
here is public\footnote{Available at \niceurl{https://data.desi.lbl.gov/public/ets/target/catalogs/} and detailed at \niceurl{https://desidatamodel.readthedocs.io}}.

\end{abstract}

\keywords{Milky Way Galaxy, Radial velocity, Stellar abundances, White dwarf stars}

\section{} 

Although DESI is conceived primarily as a cosmological experiment, in bright time the survey will obtain spectra of Milky Way stars 
within the DESI footprint selected
from the \gaia\ DR2 catalog \citep{GaiaDR2} and the DESI Legacy Imaging Surveys \citep[LS;][]{Dey2019}. This Milky Way Survey (MWS) at Galactic latitudes $\vert b \vert > 22^\circ$ will share the DESI focal plane with the Bright Galaxy Survey (BGS;
Ruiz-Macias et al. 2020), using approximately half of the 
DESI fibers available during bright time.
In addition, bright stars across the entire sky (including Galactic plane sources) will be observed in twilight and poor weather conditions.

The MWS target selection is designed to be simple, inclusive, and amenable to forward modeling. 
It will yield an essentially magnitude-limited random sample of stars over a significant fraction of the sky that can be compared 
to theoretical predictions, a philosophy similar to that of the SDSS main galaxy sample. 
Comprised of $\sim$8 million sources with \gaia\ parallaxes and proper motions but fainter than the $G=16$ limit of \gaia's radial velocity spectrograph, 
MWS will more completely sample 
stellar kinematics 
far from the Galactic center.

\subsection{MWS Main Sample}

\begin{figure}[h!]
\begin{center}
\includegraphics{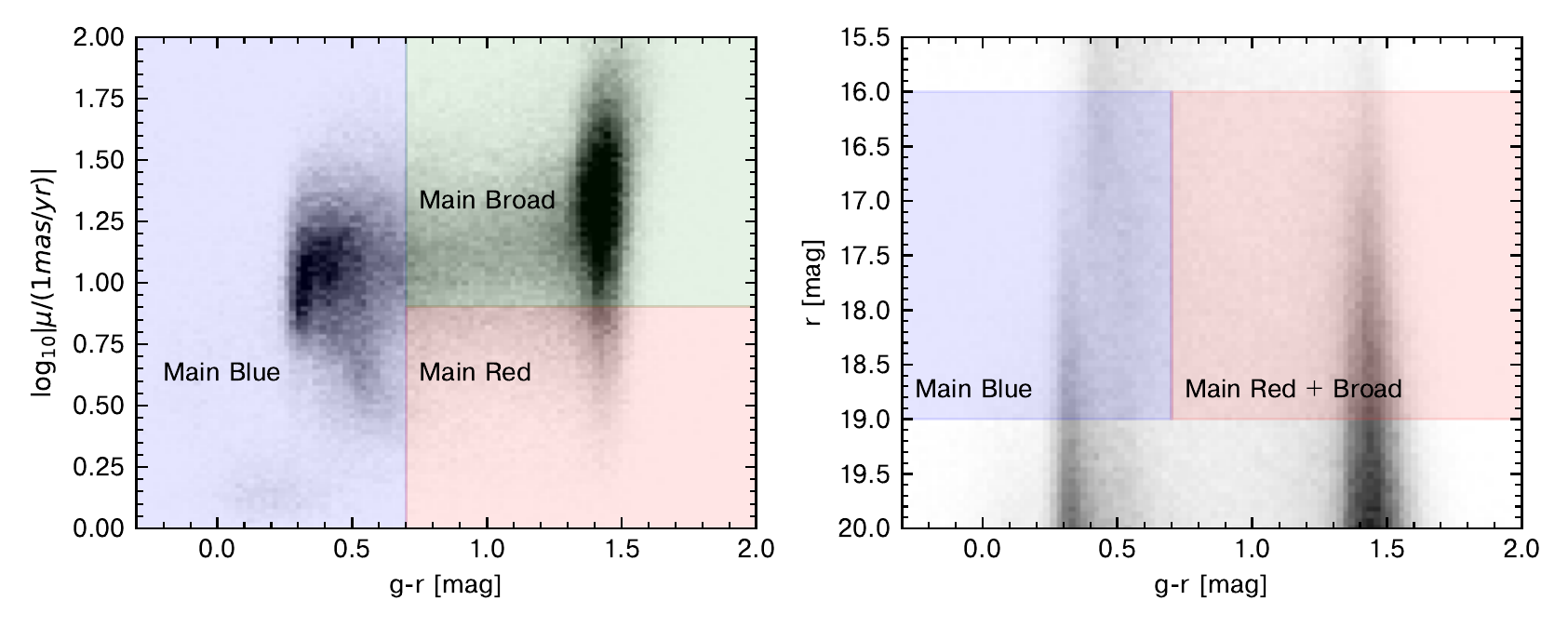}
\caption{{\it Left:} Source density as a function of
        $g-r$ color and proper motion for stars in the magnitude range  $16<r<19$ near the North Galactic Pole. 
        Redward of $g-r=0.7$, cuts in parallax
        ($\mathrm{\pi}<3\sigma_{\mathrm{\pi}}+1$) and proper motion exclude
        fast-moving nearby dwarfs.
        Main Red and Main Blue targets
        are observed at equal priority and
        constitute the bulk of the MWS main survey sample; 
        Main Broad targets
        are observed at lower priority. {\it Right:} Selection in color--magnitude space, illustrating that Main Blue targets are halo/disk main sequence turn-off (MSTO) and BHB stars, while Red/Broad targets are MW giants/dwarfs. }

\label{fig:mwsselection}
\end{center}
\end{figure}

The MWS Main sample will target all stars (i.e., sources of type ``PSF'' from LS with \gaia\ DR2 \textsc{astrometric\_excess\_noise} $< 3$~mas) in the range $16<r<19$~mag, where $r$ is the LS apparent $r$-band magnitude corrected for Galactic extinction. A small number of stars with uncorrected magnitude  $r_{\mathrm{obs}}<20$ will be excluded to ensure spectra of sufficient signal-to-noise (the DESI bright-time survey footprint is mostly at high latitude, hence low extinction). 
Stars will be assigned to one of following three target classes based on their color and \gaia\ DR2 astrometry (Fig.~\ref{fig:mwsselection}).

\begin{itemize}

\item \textbf{MWS Main Blue}:  \textbf{all} stars with  $g-r<0.7$~mag.  

\item \textbf{MWS Main Red}:  stars with  $g-r \ge 0.7$~mag; good parallaxes (Gaia \textsc{astrometric\_params\_solved = 31}, parallax $\pi < \mathrm{max}(3\sigma_{\pi}, 1)\; \mathrm{mas}$); and very small proper motion ($|\mu| < 7 \;\mathrm{mas/yr}$).

\item \textbf{MWS Main Broad}: stars with $g-r \ge 0.7$ not included in MWS Main Red.

\end{itemize}

For $g-r<0.7$, the selection is a magnitude-limited random sampling, valuable for high-resolution kinematic studies of the thick disc and nearby halo using turn-off
stars. For $g-r\ge0.7$, the Main Red astrometric criteria exclude a significant fraction of nearby dwarfs and hence increase the probability of targeting more distant halo giants, useful for studies of kinematic and stellar population gradients to $\sim100$~kpc. The Main Blue and Main Red selections are 
weighted equally in fiber assignment. 
The remaining stars are included in the Main Broad sample and assigned fibers at lower priority. 
The completeness and purity of the Main Red sample will be evaluated during the DESI Survey Validation (SV) phase (a precursor to the DESI `main' science survey). The magnitude, color, proper motion, and parallax limits will be adjusted, if necessary, before the survey commences. We will prioritize all known \gaia\ RR Lyrae with 
$14 < G < 19$ \citep{GaiaRRLyrae}, and evaluate a blue horizontal branch (BHB) star  selection based on \gaia\ and LS in this range. 

\subsection{The Backup Survey}

During poor weather and twilight conditions, when the 
MWS Main and BGS 
observations are not possible, DESI will observe an ``unbiased'' bright star sample ($10.5<G<16$~mag) at declination $\delta\ge -30^\circ$. 
This Backup Survey is expected to yield spectroscopic parameters, 
abundances, and kinematics
for  several million stars, creating a rich resource for studies of disk stellar populations, stellar evolution, and Galactic structure.

\subsection{Sparse high-value targets}

MWS will also include
sparse ($<10$ deg$^{-2}$) samples of high-value targets. These will be given higher priority for fiber assignment
but will not perturb the main sample selection function significantly. 

\paragraph{\textbf{White Dwarfs}\label{sec:mws_wd_selection}} 
White dwarf (WD) stars can constrain the local star formation history, the nature of SN\,Ia progenitors, and the composition of exo-planetesimals. Selected following \cite{GentileFusillo2019}, their priority will be the highest of all bright-time targets. 

\paragraph{\textbf{Stars within 100\,pc}} \gaia\ is complete for stars within 100~pc, and DESI will provide a comprehensive stellar population census of the solar neighborhood. This will establish strong constraints on the initial mass function and 
local chemical evolution. Target selection: $16 \le G \le 20$, $\pi + \sigma_{\pi} \ge 10$~mas. Stars with $G<16$ are in the Backup Survey. \\
 
MWS may include these additional high-value targets:

\paragraph{\textbf{Distant halo tracers}} RR Lyrae, BHB and giant stars 
are excellent tracers of halo structure because reliable distances can be determined for them. We will target faint ($G>19$) \gaia\ RR Lyrae and color-selected BHBs, and use DDO51 photometry over 10,000~deg$^2$ \citep{Slater2016} to select a sparse sample of giants. 

\paragraph{\textbf{Close white dwarf binaries}} Understanding the evolution of WD binaries, which include SN\,Ia progenitors and low-frequency gravitational wave sources, requires robust observational constraints \citep{Toloza2019}. These targets will be selected using \textit{Gaia/GALEX} data as follows: $\pi/\sigma_\pi>5$, $16 \le G \le 20$, $FUV + 5\log10(\pi[\arcsec]) + 5 > 1.5 + 1.28\times(FUV-G)$. 

\paragraph{\textbf{Cluster members}} We will use proper motion to select likely members of a few bright dwarf galaxies and globular clusters out to $\sim10$ times their half-light radius. DESI is ideal for surveying the environs of such objects for signs of tidal interactions and extended halos. 

\acknowledgments

APC is supported by the Taiwan Ministry of Education Yushan Fellowship and MOST grant 109-2112-M-007-011-MY3. SK was partially supported by NSF grants AST-1813881, AST-1909584. TSL is supported by NASA through Hubble Fellowship grant HST-HF2-51439.001. MYW acknowledges the support of the McWilliams Postdoctoral Fellowship. MV is supported by NASA-ATP awards NNX15AK79G and 80NSSC20K0509 and the Michigan Institute for Computational Discovery and Engineering (MICDE).


This research is supported by the Director, Office of Science, Office of High Energy Physics of the DOE under Contract No. DE–AC02–05CH1123, and by NERSC, a DOE Office of Science User Facility under the same  contract; additional support for DESI is provided by NSF, Division of Astronomical Sciences under Contract No. AST-0950945 to the NSF’s NOIRLab; the UK STFC; the Gordon and Betty Moore Foundation; the Heising-Simons Foundation; the French Alternative Energies and Atomic Energy Commission (CEA); the National Council of Science and Technology of Mexico; the Ministry of Economy of Spain, and by the DESI Member Institutions.  The authors are honored to be permitted to conduct astronomical research on Iolkam Du’ag (Kitt Peak), a mountain with particular significance to the Tohono O’odham Nation.  

\bibliography{biblio}

	\end{document}